# Estimations of Cross-Sections for Photonuclear Reaction on Calcium Isotopes by Artificial Neural Networks

S. Akkoyun[a,b], H. Kaya[a]

[a]Department of Physics, Sivas Cumhuriyet University, Sivas, 58140, Turkey

[b]Articial Intelligence Systems and Data Science Application and Research Center, Sivas Cumhuriyet University, Sivas, 58140, Turkey

**Abstract**

The nuclear reaction induced by photon is one of the important tools in the investigation of atomic nuclei. In the reaction, a target material is bombarded by photons with high-energies in the range of gamma-ray energy range. In the bombarding process, the photons can statistically be absorbed by a nucleus in the target material. Then the excited nucleus can decay by emitting proton, neutron, alpha and light particles or photons. By performing photonuclear reaction on the target, it can be easily investigated low-lying excited states of the nuclei. In the present work, ($\gamma$, n) photonuclear reaction cross-sections on different calcium isotopes have been estimated by using artificial neural network method. The method is a mathematical model that mimics the brain functionality of the creatures. The correlation coefficient values of the method for both training and test phases being 0.99 indicate that the method is very suitable for this purpose.

## 1. INTRODUCTION

In the experimental nuclear physics studies, reactions induced by photons are one of the important tools. In these types of reactions, the target nuclei are bombarded by high-energy photons and the photons can be absorbed by a nucleus in the target material. Because of a nuclear process can be observed in the reaction, these are called as photonuclear reaction [1]. In order to release excess energy by the nucleus, one of the decay process is governed via emitting proton, neutron, alpha and light particles or photons. In the case of neutron emission, the reaction is called as photo-neutron reaction. By emitting particles from the target nucleus, a stable or unstable isotope can be formed. The unstable isotope goes to stable one by beta decays. The half-life of the radioisotopes can also be determined in these types of reactions.

The cross-section values for photo-neutron reactions for different isotopes and energies are determined either experimentally or theoretically [2, 3]. One of the most used theoretical codes for this purpose is TALYS computer code [4]. TENDL database [5] is based on this code and other sources such as ENDF [6]. The code is a system for the analysis and prediction of nuclear reactions. The basic objective behind its construction is the simulation of nuclear reactions that involve neutrons, photons, protons, deuterons, tritons, $^3$He- and alpha-particles, in the 1 keV - 200 MeV energy range and for target nuclides of mass 12 and heavier. To achieve this, it is implemented a suite of nuclear reaction models into a single code system.

Calcium (Ca) is a metal with a silver color whose structure is cubic crystal [7]. It has 25 known isotopes and 5 of them are stable. The abundances of the isotopes in percentage are 96.941, 0.647, 0.135, 2.086 and 0.187 for $^{40}$Ca, $^{42}$Ca, $^{43}$Ca, $^{44}$Ca and $^{46}$Ca, respectively. Ca has also long-lived unstable isotopes which are suitable for target material which are $^{41}$Ca and $^{45}$Ca isotopes with 99400 years and 162.6 days half-lives. Ca element is one of the fundamental target materials for experimental nuclear structure studies. It also plays a number of biologically important roles that have been explored for a long time with various techniques available for medical science [8]. One of the most important of these techniques is investigation of calcium functions in the body with the assist of a radioactive isotope. Apart from the stable isotopes, the radioactive Ca isotopes have been artificially produced which are possible radioactive tracers.

The one of the way to produce the radioactive Ca isotopes is photo-neutron (γ, n) reaction. $^{41}$Ca, $^{45}$Ca and $^{47}$Ca can be generated by using photo-neutron reactions performed on $^{42}$Ca, $^{46}$Ca and $^{48}$Ca stable isotopes. Therefore, the information about the cross-sections on Ca according to different energy values for these reactions is important. In the present study, artificial neural network (ANN) method [9] has been used for the prediction of (γ, n) reaction cross sections in different energies from threshold energy values to 200 MeV on stable or long-lived Ca isotopes. The data are taken from TENDL2019 library [5]. ANN is a machine learning tool as a mathematical model that mimics brain functionality which is composed of layers including neurons in each. The method generates its own output as close as the desired values. One of the advantages of the method is it does not need any relationship between input and output data variables. Another advantage of the method is that in case of missing data, it can complete missing data thanks to its learning ability. Recently, ANN has been used in many fields in nuclear physics. Among them the studies performed by our group are developing nuclear mass systematic [10], obtaining fission barrier heights [11], obtaining nuclear charge radii [12], estimation of beta decay energies [13], approximation to the cross sections of Z boson [14, 15], determination of gamma-ray angular distributions [16] and estimations of radiation yields for electrons in absorbers [17].

## 2. MATERIAL and METHODS

ANN (artificial neural network) is a very powerful mathematical tool that is used when standard techniques fail [9]. The method mimics brain functionality and nervous system. It is composed of three different layers which are input, hidden and output layers. Each layer has its own neurons. The neurons are processing units. The neurons in a layer are connected to the neurons only in the next layer by adaptive synaptic weights. The input neurons receive the data which are independent variables of the problem. The data is transmitted to the hidden layer neurons by multiplying the weight values of the connections. The all data entering the neurons are summed and the summed net data are activated by appropriate functions. The hidden neuron activation function can be theoretically any well behaved nonlinear function. In this study, a sigmoid-like function (tangent hyperbolic) has been used for the activation. Finally, the data is transmitted to the output layer neurons and predictions have been done for the dependent variables. Because of the layered structure, a particular type of ANN is called layered ANN. In Fig.1, we have shown the 3-20-20-1 ANN structure which is used in this study for the prediction of the reaction cross sections.

The inputs were neutron number (N) of the target, mass number (A) of the target and photon energy (E) impinging upon the target. $^{40}$Ca, $^{41}$Ca,

$^{42}$Ca, $^{43}$Ca, $^{44}$Ca, $^{45}$Ca, $^{46}$Ca and $^{48}$Ca isotopes are considered in the target. The desired output was photo-neutron reaction cross-section for these different Ca isotopes. There is no rule for the determination of the hidden layer and neuron numbers. It depends of problem nature and determined after several trials. In this work, two hidden layer with 20 neurons in each were chosen as 2 and 4 in each, respectively.

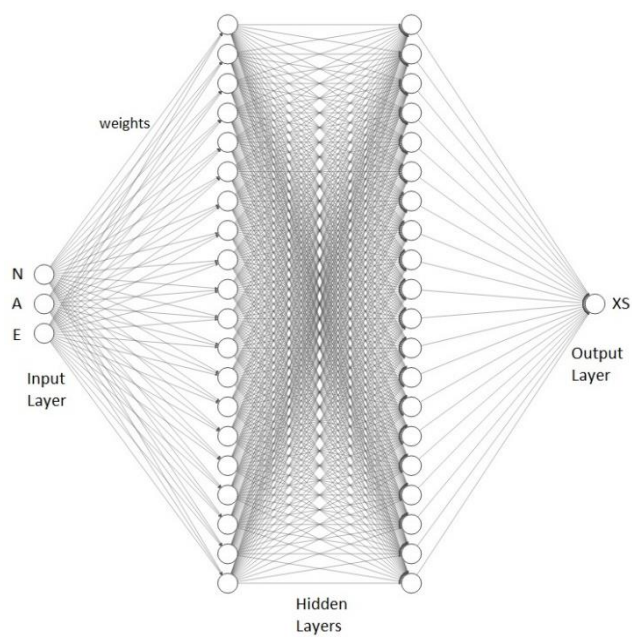

Figure 1. ANN with 3-20-20-1 structure for the prediction of photo-neutron cross sections performed on Ca isotopes

The main goal of the method is the determination of the final weight values between neurons by starting random values. The ANN with best weights can give the ANN outputs as close as to the desired values. ANN is two step processes. In the first, ANN is trained (training) for the determination of the final best weights by given input and output data values. By the appropriate modifications of the weights, ANN modifies its weights until an acceptable error level between ANN and desired outputs. The error function was mean square error (MSE) in this study. MSE gives the average of the squares of the difference between the desired and the neural network output values. Correlation coefficient (R) takes values between -1 and 1. It gives the correlation between values. Over-training and network memorization can be prevented by validation dataset which is not included in training. If validation is not made, it is not noticeable that the network is trained or memorized during the training stage.

In the second step (test), another dataset of the problem is given to ANN and the results are predicted by using the final weights. If the predictions of the test data are good, the ANN is considered to have learned the relationship between input and output data.

In this work, neural network tool of MATLAB program [18] was used for the estimations. The data was divided into three separate sets for training (70%), validation (15%) and test (15%) stages. The whole data were obtained from TENDL database. In the training step, Levenberg–Marquardt [19, 20] backpropagation algorithm was used.

## 3. RESULTS and DISCUSSION

A total of 282 data were used for calculations, 198 of which are for training and 42 for the validation of ANN and 42 for the test of ANN. The data for (γ, n) reaction cross-sections in the literature are studied from threshold energy values to 200 MeV. After the determination of the final weights, the ANN is first used in the training and validation datasets. As can be seen in regression plots (Figure 2) that the R values are 0.99998 and 0.99827 for the training and validation data. All data concentrated on the X = Y line indicates that the method is very suitable for this purpose.

For the test dataset R values is still high by getting values 0.9988. MSE, minimum and maximum errors for the all data are 0.202 mb, $10^{-5}$ mb and -3.27 mb, respectively. For the all data the R values is 0.99941. The R values obtained for each datasets which are close to 1 show the strength of the relationship. However, the R value in the test stage is lower than the training stage. This indicates that ANN does not memorize data, as can be seen also in the validation results. In the training stage, the MSE value gets its minimum value for the validation data in 43th epoch of total 1000. We have shown in Figure 3 that the plot for validation starts to increase after this epoch with the value of 0.66641. The final MSE value of the training is in the order of $10^{-5}$.

In Figure 4, we have given the photo-neutron reaction cross section values for Ca isotopes from ANN method in comparison with the available literature data. As is clear in figure that the ANN estimations are in harmony with the literature data. The eight peaks belong to $^{40}$Ca, $^{41}$Ca, $^{42}$Ca, $^{43}$Ca, $^{44}$Ca, $^{45}$Ca, $^{46}$Ca and $^{48}$Ca isotopes, respectively. For $^{40}$Ca isotope, the reaction cross-section values are the lowest among the other isotopes for all energy values in the energy range. For $^{41}$Ca isotope, the cross-sections are relatively lower than the other higher Ca isotopes. The larger cross-section values are for $^{44}$Ca and $^{45}$Ca isotopes.

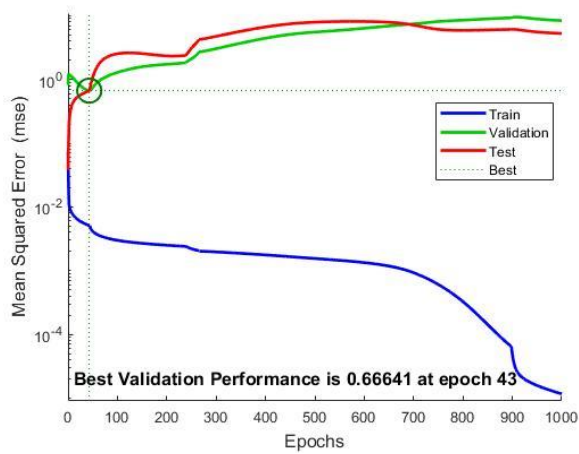

Figure 3. The performance plots of the ANN for training, validation and test datasets.

Also, it can be seen in Figure 4 that the maximum cross-section values are 2.98 mb in 22 MeV for $^{40}$Ca, 17.22 mb in 19 MeV for $^{41}$Ca, 33.82 mb in 20 MeV for $^{42}$Ca, 47.51 mb in 19 MeV for $^{43}$Ca, 59.52 mb in 19 MeV for $^{44}$Ca, 60.38 mb in 19 MeV for $^{45}$Ca, 56.90 mb in 18 MeV for $^{46}$Ca and 40.19 mb in 18 MeV for $^{48}$Ca. The cross-sections get its maximums about 18-20 MeV in the investigated energy range of threshold to 200 MeV. The reaction thresholds are 16, 9, 12, 8, 12, 8, 11 and 10 MeV, respectively, for to $^{40}$Ca, $^{41}$Ca, $^{42}$Ca, $^{43}$Ca, $^{44}$Ca, $^{45}$Ca, $^{46}$Ca and $^{48}$Ca isotopes.

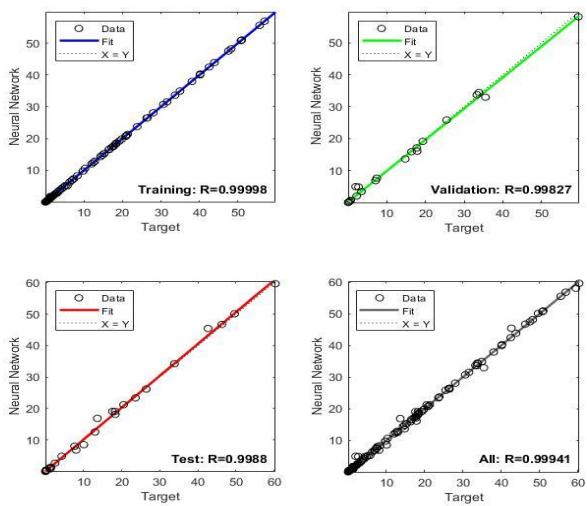

Figure 2. The comparisons of the ANN results by target values for training (upper-left), validation (upper right), test (lower-left) and all (lower-right) datasets.

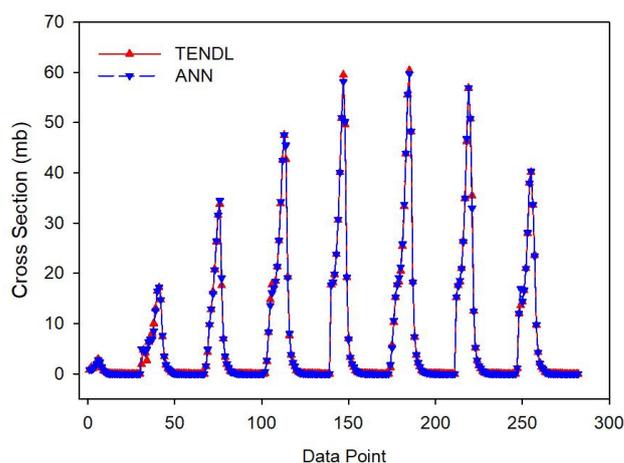

Figure 4. Photo-neutron reaction cross-sections performed on Ca isotopes from ANN estimation (ANN) and the literature data (TENDL).

## 4. CONCLUSIONS

In this work, (γ, n) photo-neutron reaction cross-sections of stable and long-lived Ca isotopes have been predicted by using artificial neural network (ANN) method in the threshold to 200 MeV energy range. The data for the applications of the ANN method have been borrowed from TENDL-2019 nuclear data library. According to the results, the ANN predictions for the cross-sections are very close to the available literature data. Therefore, one can use ANN method for the obtaining of photonuclear reaction cross-sections whose values are not available in the literature.

### 4.1. Acknowledgments

This study is supported by Sivas Cumhuiyet University Scientific Research Projects Coordination Unit. Project Number: F-616.